# Observation of Hermitian and Non-Hermitian Diabolic Points and Exceptional Rings in Parity-Time symmetric ZRC and RLC Dimers

Stéphane Boris. TABEU, Fernande Fotsa-Ngaffo and Aurélien Kenfack-Jiotsa

*Abstract*—We present the observation of diabolic points in Hermitian and non-Hermitian electronics dimers. The condition of unbreakable Parity-time symmetry is established for both PT-symmetric ZRC and RLC dimers. We show how appears non-Hermitian degeneracy points in the spectrum and how they are protected against a Hermitian perturbation. When a non-Hermitian perturbation is added in the setup, the non-Hermitian diabolic point (NHDP) turns into a ring of exceptional points as in some Dirac and Weyl semimetals. Some experimental simulations of oscillations around these particular points in LTspice are in perfect accordance with the one predicted analytically and numerically. This work opens a gold road for investigations on topological electrical circuits for robust transport of information at room temperature.

## I. Introduction

THE race for the development of novel components for robust transport of information is now under way in many main fields of Physics. Beginning on Condensed matter in material Science, passing through Photonics, Optics, Atomic and Molecular Physics, Quantum Topology, Mechanics, Plasmonics, Metamaterials and Topoelectrical Electronics, all these axes of research are activated. The properties of the Parity-Time (PT) symmetry play a great role in the design of the systems used since Carl Bender in 1998 demonstrated that a non-Hermitian Hamiltonian (NHH) may have an entirely real spectrum for a well-defined potential [1-4]. In Parity-Time symmetric systems, there is a parameter related to the gain and loss which controls the degree of hermiticity. When the eigenvalues of the non-Hermitian Hamiltonian are not more the ones of the PT-symmetry, the system enters into the broken phase. The passage from the exact phase where the PT-symmetry and the non-Hermitian Hamiltonian share the same set of eigenvalues and the broken phase is guaranteed by an exceptional point EP [1-29]. Generally, at EPs, the eigenvalues and the eigenvectors coalesce differently to what is observed at accidental degeneracy called diabolic point (DP) in Hermitian systems where only the eigenvalues coalesce when the eigenvectors are orthogonal [8, 16, 18-21]. Nowadays, EPs play a major role in Physics. It occurs for example at this point: nonreciprocal properties such as unidirectional invisibility and reflectionless; laser self-termination, enhancement of sensing in Optics, Photonics and telemetry [5, 7, 18-30]; exhibition of the spin–orbit locking and chiral modes just to name a few [31-59]. The study of EPs remains a wide-open field of research. The results permit the classification of systems according to the type of coalescing of eigenvalues. Then we can have the classes of Weyl or Dirac systems in Material Science, in Optics, in Photonics, in Cold Atoms, in Magnon Polaritons, superconducting qubits and in Electronics [31-59]. Usually, Weyl points occurs in pair in systems such that such that the total charge in the Brillouin zone remains zero. In 2016, Qian Lin et al show how to achieve four Weyl points in a photonic system [36]. In 2019, Yuehui Lu et al. demonstrated also the presence of four Weyl points in the first Brillouin zone with a Weyl circuit analogous to those of cold atoms [37]. The same year, Alexander Cerjan et al proved that one can realize a single Weyl point (WP) in photonics crystals as in acoustic phonons in the center of the Brillouin zone [40]. In Topoelectric systems, the realization of the necessary $\pi$-flux for the appearance of the Weyl point is not an easier task due the interferences produced by the proximity of inductors in the setup [45]. The existing system with inductors and capacitors are in second order resonance with a natural frequency $\omega = \sqrt{1/LC}$. More recently, Tabeu *et al* in [8, 16, 60] demonstrated how to build electronic oscillators (ZRC and ZRL) of first order resonance with positive and negative frequencies using imaginary resistors $Z = jR$. In ZRC cell for example, the natural frequency is $\omega = 1/rC$ and its sign is function of the capacitor and the imaginary resistor used. It has been demonstrated in literature

Stéphane B. Tabeu is with the Laboratory of Electronics and Signal Processing, Department of Electrical and Telecommunication Engineering, National Advanced School of Engineering, University of Yaoundé I - P.O. Box 8390, Yaoundé, Cameroon, the Laboratory of Mechanics, Materials and Structures, Department of Physics, University of Yaoundé I - P.O. Box 812, Yaoundé, Cameroon and Nonlinear Physics and Complex Systems Group, Department of Physics, The Higher Teacher's Training College, University of Yaoundé I - P.O. Box 47, Yaoundé, Cameroon (stephaneboris@yahoo.fr).

Fernande Fotsa-Ngaffo is with the Institute of Wood Technologies, University of Yaoundé I - P.O. Box 306, Mbalmayo, Cameroon and Department of Physics, Faculty of Science, University of Buea, P.O. Box 63, Buea, Cameroon (ngafotsa@yahoo.com).

Aurélien Kenfack-Jiotsa is with the Nonlinear Physics and Complex Systems Group, Department of Physics, The Higher Teacher's Training College, University of Yaoundé I - P.O. Box 47, Yaoundé, Cameroon and High-Tech and Slow Technology (HITASTEC), P.O. Box 105, Yaoundé, Cameroon (kenfack@yahoo.com)..

that a PT-symmetric system may exist without phase transition and exceptional points. These features resulted from the presence of the nonlinearity or not. In the linear Electronics, such a system may have negative capacitors (ZRC) or negative inductors (ZRL) with gyrators.

We aim to present in this work the presence of Hermitian diabolic points (HDPs), non-Hermitian diabolic points (NHDPs) and ring of exceptional points (REPs) in the linear unbreakable PT-symmetry with ZRC and RLC PT-dimers coupled by gyrators, capacitors, inductors or imaginary resistors.

The work is organized as follows: In Section II, the models of linear unbreakable PT-symmetric dimers are presented. In Section III, is presented the system of first order resonance named the ZRC PT-dimer with its properties. In Section IV is presented the second order resonance RLC PT-dimer with its properties. Finally, we give a conclusion and some perspectives for future works.

## II. LINEAR UNBREAKABLE PT-SYMMETRY

### A. The general model of Parity-time symmetric two-level systems for non-Hermitian quantum mechanics

We consider the generic Hamiltonian $H$ in the classical Schrodinger equation in (1) such that $H$ is related to the all the matrices of Pauli.

$$j \frac{d}{dt}|\psi(t)\rangle = H|\psi(t)\rangle \quad (1)$$

where: $j$ is the imaginary unit $|\psi(t)\rangle = (\psi_0, \psi_1)^T$; $H = h_0 \mathbf{I} + h_x \sigma_x + h_y \sigma_y + h_z \sigma_z$. $\mathbf{I}$ is the $2\times 2$ matrix unit and $\sigma_x$, $\sigma_y$ and $\sigma_z$ are Pauli's matrices.

$$\sigma_x = \begin{pmatrix} 0 & 1 \\ 1 & 0 \end{pmatrix}; \ \sigma_y = \begin{pmatrix} 0 & -j \\ j & 0 \end{pmatrix}; \ \sigma_z = \begin{pmatrix} 1 & 0 \\ 0 & -1 \end{pmatrix} \quad (2)$$

$h_x$, $h_y$ and $h_z$ are complex coefficients function of the independent real parameters such that the eigenvalues of the Hamiltonian resulted from $\det(H - \omega \mathbf{I}) = 0$ are given as follow:

$$\omega_{\pm} = h_0 \pm \sqrt{h_x^2 + h_y^2 + h_z^2} \quad (3)$$

If all the coefficients are real, the system is Hermitian. The system is PT-symmetric if (4) is satisfied.

$$[PT, H] = 0 \quad (4)$$

where $P = \sigma_x$ is the Parity operator and $T = \mathcal{K}$ is the time reversal operator with $\mathcal{K}$ being the complex conjugation operation. That is $h_0$, $h_x$ and $h_y$ must be real whereas $h_z$ must be purely imaginary. Looking at (3), the eigenvalues are real when $h_x^2 + h_y^2 > -h_z^2$ and complex in the contrary case. When $h_x^2 + h_y^2 = -h_z^2$, we are at the exceptional point where the square roof of (3) is cancelled and the eigenfrequencies coalesce such that $\omega_{\pm} = h_0$.

### B. Unbreakable Parity-Time Symmetry.

Now, we consider the Hamiltonian $H$ such that $h_0(X,Y,Z,...)$, $h_x(X,Y,Z,...)$ and $h_y(X,Y,Z,...)$ are real coefficients and $h_z(X,Y,Z,...)$ is a purely imaginary coefficient, all function of the independent real parameters $X$, $Y$, $Z$ and others. We choose new parameters $k_0$ $k_x$, $k_y$ and $k_z$ with $X \neq -\frac{1}{2}$ such that:

$$h_x = \frac{k_x}{\sqrt{1+2X}} \ ; \ h_y = \frac{k_y}{\sqrt{1+2X}} \ ; \ h_z = \frac{k_z}{\sqrt{1+2X}} \quad (5)$$

The eigenvalues expressed in (3) take the form:

$$\omega_{\pm} = \frac{k_0 \sqrt{1+2X} \pm \sqrt{(1+2X)(k_x^2 + k_y^2 + k_z^2)}}{1+2X} \quad (6)$$

The possibility of having a real spectrum is function of the combination of $k_0$ $k_x$, $k_y$ and $k_z$ and the behavior of $\sqrt{1+2X}$ according to the values of $X$. Two major possibilities are raised to have an unbreakable Parity-Time symmetry with real values of eigenfrequencies with (4) satisfied:

➢ $X > -1/2$ and $k_x^2 + k_y^2 + k_z^2 > 0$ for all values of $Y$, $Z$ and others. The others values of $X$ are those of breakable PT-symmetry.

➢ $X < -1/2$ and $k_x^2 + k_y^2 + k_z^2 < 0$ for all values of $Y$, $Z$ and others. The others values of $X$ are also those of breakable PT-symmetry.

The eigenvalues values have a asymptotic behavior when approaching $X = -\frac{1}{2}$ which marks in both cases the transition between breakable PT-symmetry and unbreakable PT-symmetry. The exploitation of the surrounding of this singular point can lead on the generation of very high frequencies for future applications in Optics, in Photonics and in Telecommunications with the avenue of 5G, 6G and next generations of networks.

## III. ZRC PARITY-TIME SYMMETRIC DIMER

### A. The ZRC model and equations of dynamics

The ZRC model is built by two active ZRC-cells coupled by a capacitor $C'$, an imaginary resistor $z = jr'$, a gyrator $G$ or two (or all) of them [8, 16]. Each cell contains in parallel a real resistor $R_n$, a capacitor $C_n$ and an imaginary resistor $Z_n$.

Their values may be positive or not without any restriction. The setup of the ZRC model is presented in Fig.1.
By applying the Kirchoff's laws at nodes 0 and 1, we have:

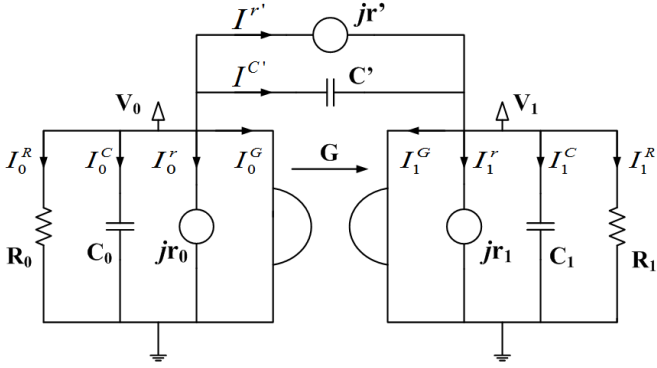

Fig. 1. The setup of the ZRC dimer. The two ZRC-cells are coupled by a capacitor $C'$, an imaginary resistor $z = jr'$, a gyrator $G$ or by two (or by all) of them. Each cell contains a resistor $R_n$, a capacitor $C_n$ and an imaginary resistor $Z_n = jr_n$.

$$\begin{cases} I_0^R + I_0^C + I_0^Z + I_0^G + I^{C'} + I^{r'} = 0 \\ I_1^R + I_1^C + I_1^Z + I_1^G - I^{C'} - I^{r'} = 0 \end{cases} \quad (7)$$

with $\begin{pmatrix} I_0^G \\ I_1^G \end{pmatrix} = \begin{pmatrix} 0 & G \\ -G & 0 \end{pmatrix} \begin{pmatrix} V_0 \\ V_1 \end{pmatrix}$ for the gyrator and $I_n^R = \dfrac{V_n}{R_n}$ ;

$I_n^C = C_n \dfrac{d}{dt} V_n$ ; $I^{C'} = C' \dfrac{d}{dt}(V_0 - V_1)$ ; $I_n^Z = -j\dfrac{V_n}{r_n}$ ;

$I^{R'} = \dfrac{1}{R'}(V_0 - V_1)$ ; $I^{r'} = -j\dfrac{1}{r'}(V_0 - V_1)$ for all others components. These considerations lead on the general form of first order ordinary differential equations:

$$\begin{cases} \dfrac{dV_0}{d\tau} = \dfrac{1}{\Delta}\left[(1+c_1)\left(j(1+v_0) - \gamma_0\right) - c_0\Gamma(g_1 + jv_1)\right]V_0 \\ \quad + \dfrac{1}{\Delta}\left[-(1+c_1)(g_0 + jv_0) + c_0\Gamma\left(j(1+v_1) - \gamma_1\right)\right]V_1 \\ \dfrac{dV_1}{d\tau} = \dfrac{1}{\Delta}\left[-(1+c_0)\Gamma(g_1 + jv_1) + c_1\left(j(1+v_0) - \gamma_0\right)\right]V_0 \\ \quad + \dfrac{1}{\Delta}\left[(1+c_0)\Gamma\left(j(1+v_1) - \gamma_1\right) - c_1(g_0 + jv_0)\right]V_1 \end{cases} \quad (8)$$

where $\tau = \omega_0 t$ ; $\omega_n = \dfrac{1}{r_n C_n}$ ; $\Gamma = \dfrac{\omega_1}{\omega_0}$ ; $c_n = \dfrac{C'}{C_n}$ ; $v_n = \dfrac{r_n}{r'}$ ;

$\kappa_n = \dfrac{r_n}{R'}$ ; $\gamma_n = \dfrac{r_n}{R_n}$ ; $g_n = r_n G_n$ and $\Delta = 1 + \sum_{n=0}^{1} c_n$.

Equation (8) is PT- symmetric when the serval conditions of (9) are satisfied :

$$\begin{cases} \Gamma = 1 \; ; \; c_0 = c_1 = c \; ; \; v_0 = v_1 = v \\ \gamma_0 = -\gamma_1 = \gamma \; ; \; g_0 = -g_1 = g \end{cases} \quad (9)$$

Hence, we can derive the effective Hamiltonian $H_{eff}$ such that (8) is transformed into (10).

$$j\dfrac{d}{d\tau}|\Psi(\tau)\rangle = H_{eff}|\Psi(\tau)\rangle \quad (10)$$

where : $H_{eff} = h_0 \mathbb{I} + h_x \sigma_x + h_y \sigma_y + h_z \sigma_z$ ; $j$ is the imaginary unit $(j^2 = -1)$, $|\Psi(\tau)\rangle = (V_0(\tau), V_1(\tau))^T$ and $\mathbb{I}$ is the $2\times 2$ unity matrix.

The coefficient are then deduced as :

$$\begin{cases} h_0 = -(1+c+v)/\Delta \; ; \; h_x = (v-c)/\Delta \\ h_y = ((1+c)g - c\gamma)/\Delta \; ; \; h_z = j(cg - (1+c)\gamma)/\Delta \end{cases} \quad (11)$$

Following (4), (5) and (9), the eigenvalues of the effective Hamiltonian are :

$$\omega_\pm = \dfrac{-(1+c+v) \pm \sqrt{(1+2c)(\gamma_{PT}^2 - \gamma^2 + g^2)}}{1+2c} \quad (12)$$

TABLE I
UNBREAKABLE PARITY-TIME SYMMETRY WITH ZRC-DIMER

| NON-HERMITIAN PARAMETER | CAPACITOR | |
|---|---|---|
| | $c < -1/2$ | $c > -1/2$ |
| Gain-loss Parameter $(\gamma \neq 0 \; ; \; g = 0)$ | UNBREAKABLE PT-SYMMETRY (WITHOUT EPs) | BREAKABLE PT-SYMMETRY (WITH EPs) |
| Gyroscopic coupling parameter $(\gamma = 0 \; ; \; g \neq 0)$ | BREAKABLE PT-SYMMETRY (WITH EPs) | UNBREAKABLE PT-SYMMETRY (WITHOUT EPs) |

with $\gamma_{PT} = \pm\dfrac{|c-v|}{\sqrt{1+2c}}$.

The different configurations of the occurrence of the unbreakable PT-symmetry are resumed in the Table I.

*B. Hermitian and non-Hermitian diabolic points, and rings of exceptional points in ZRC-Dimer*

We will focus our attention on the behavior of the system in the unbreakable regions with gain/loss parameter or with the gyroscopic coupling.

*Hermitian point in PT-symmetric ZRC-dimer*

In the setup parameter, the case $c = -1$ without the gyroscopic coupling is mostly important due to the fact that the coupling induced Hermitian behavior whereas it is a balanced gain and loss in the systems. The coefficient $h_0$, $h_x$, $h_y$ are always real and $h_z = 0$. Consequently, the effective Hamiltonian $H_{eff}$ is Hermitian $H_{eff}^\dagger = H_{eff}$. The new form of the Hamiltonian is given at (13)

$$H_{eff} = \begin{pmatrix} v & -(1+v-j\gamma) \\ -(1+v+j\gamma) & v \end{pmatrix} \quad (13)$$

The diabolic point occurs when $\gamma = 0$ and as aforementioned, the eigenvalues coalesce and the eigenvectors are orthogonal. For an initial input, there is not amplification nor attenuation. The frequency oscillations from (12) are tuned by the amounts of the imaginary coupling and gain loss parameter. The real part of the voltage for the case $v = 0$ is presented in fig (2).

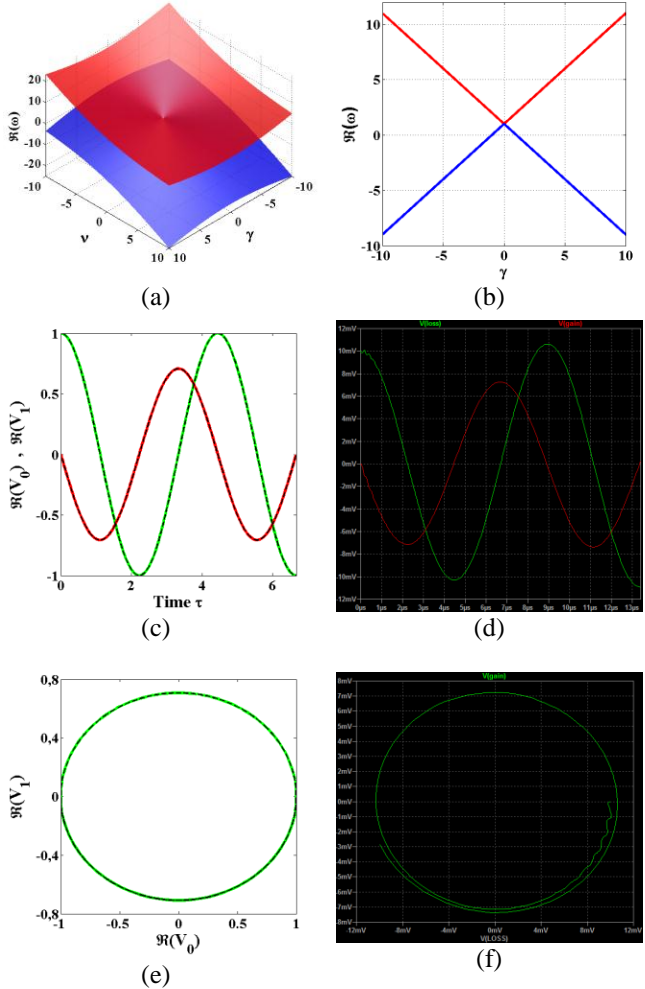

Fig. 2: (a) The Hermitian diabolic point at $c = -1$ in unbreakable PT-symmetry. With this amount of capacitive coupling, the system is Hermitian however possessing gain and loss. (b) The 2D-plot of the diabolic point at $v = -1$. (c) the real part of the real voltage across capacitors of the setup. (e) the Lissajous' curve of the real voltages across capacitors of the setup. $(c = -1; \gamma = 1; v = 0; g = 0)$. (d) and (f) Experimental simulations in LTspice at $f \approx 225 \, kHz$. $(C = 1nF; C' = -1nF; Z = j2k\Omega; R = 2k\Omega)$

This setup may be very useful in the miniaturization of antennas, in novel devices for optoelectronics and quantum electronics. With the adding of the nonlinearity one can perform robust circuit for the mimic at room temperature of the behavior of the self-trapping or the Bose-Einstein condensate.

*Non-Hermitian diabolic point in PT-symmetric ZRC-Dimer*

The diabolic point is a suitable feature more prized in condensed matter systems. His achievement is sometimes difficult due to the lack of judicious management of loss, gain or the defaults in structure. It has been proved that a material or waveguides, possessing particular point like Weyl points may be robust in the transport of information according to its topological properties.

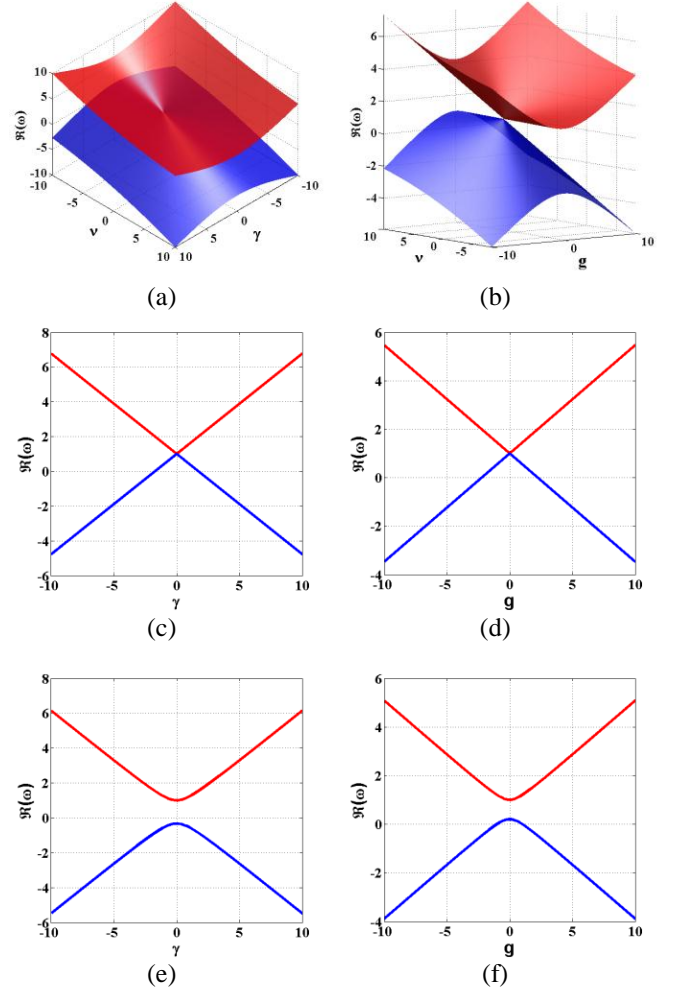

Fig. 3: (a) The non-Hermitian diabolic point at $c = -2$ in unbreakable PT-symmetry induced by the gain and loss. (b) The non-Hermitian diabolic point at $c = 2$ in unbreakable PT-symmetry induced by gyroscopic coupling. (c) The 2D-plot of the NHDP at $v = -2$. (d) The 2D-plot of the NHDP at $v = 2$. (e) and (f) The splitting of the eigenvalues when $c = v$ is no more respected respectively for the gain/loss $(\gamma \neq 0; g = 0)$ case $c = -2$ and the gyroscopic coupling $(g \neq 0; \gamma = 0)$ case $c = 2$. The splittings in both cases happen for $v = 0$.

In Photonics for example, Cerjan demonstrated experimentally how to isolate a single Weyl point in a bipartite array of helical waveguides with the same orientation [40]. We present in this work the occurrence of the NHDP in the unbreakable PT-symmetric electronic dimer. It happens at the degeneracy of the eigenvalues of system where the spectrum is gapless. It obeys to the condition $c = v$ which is similar but different to Thresholdless PT-transition in the breakable PT-symmetry [7-8, 16]. A Hermitian perturbation changes only the location of the NHDP in the space of parameters. As depicted in Table I, NHDP occurs at $c < -1/2$ when the presence of the non-hermiticity in the setup is assured only by the gain/loss. In the case of $c > -1/2$, the non-hermiticity is assured by the gyroscopic coupling. In all theses cases, the system remains exclusively in the exact phase of the PT-symmetry. In Figs. 3 and 4 are presented the NHDP point for both cases in $c \neq -1/2$ and the experimental simulation in LTspice of oscillations in the setup respectively.

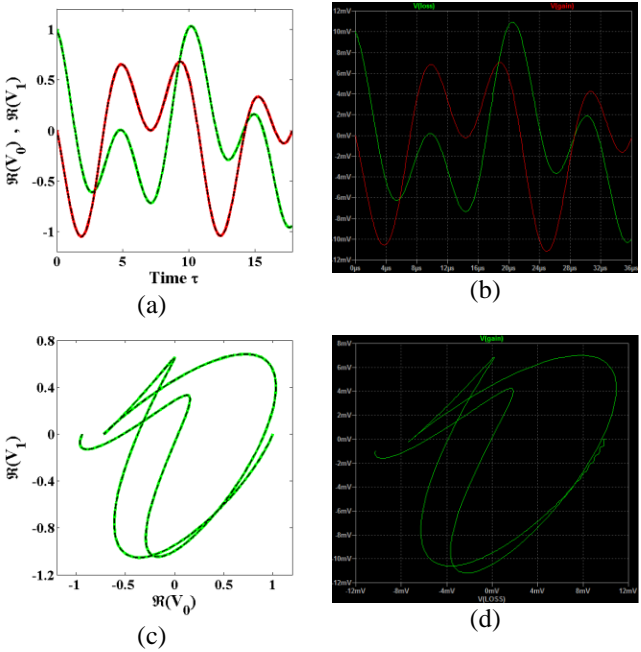

Fig. 4: (a) The real parts of the voltage across capacitors of the setup . (b) the Lissajous' curve of the real parts of voltages across capacitors of the setup. $(c = -2; \gamma = 1; \nu = 0; g = 0)$. (c) and (d) Experimental simulations in LTspice at $f \approx 140 kHz$ $(C = 1nF; C' = -2nF; Z = j2k\Omega; R = 2k\Omega)$

*Rings of exceptional points in PT-symmetric ZRC-Dimer*

When both the parameters responsible of the non-hermiticity are present in the system, the Hermitian and non-Hermitian diabolic point are transformed into a REP. Concretely, the gyroscopic coupling restores the exceptional points on a nodal line in the unbreakable PT-symmetry induced by gain/loss and vice-versa. In both cases, the nodal line of exceptional points follows the equilateral hyperbolae depicted in (14).

$$\gamma^2 - g^2 = \gamma_{PT}^2 \quad (14)$$

At Fig 5, it is presented the NHDPs in the two intervals of values of the capacitive coupling $c < -1/2$ and $c > -1/2$. One can easily observed the appearance of the non-zero imaginary part with symbolize the broken phase of the PT-symmetry.

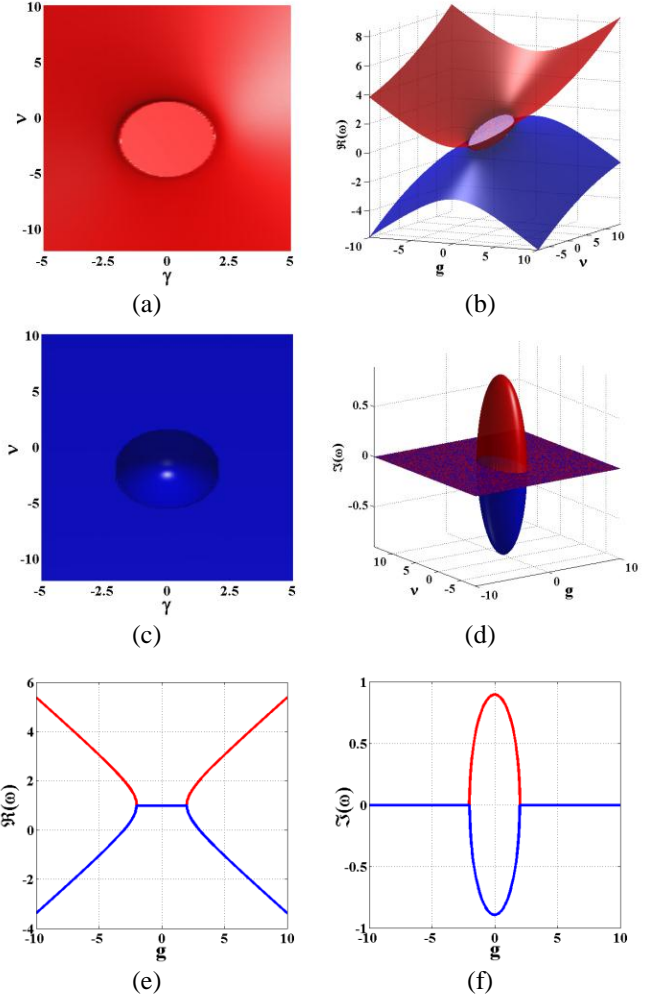

Fig. 5: (a) The ring of exceptional points at $c = -2$ in unbreakable PT-symmetry induced by the gain and loss. (b) The ring of exceptional points at $c = 2$ in unbreakable PT-symmetry induced by gyroscopic. (c) and (d) The imaginary of the eigenvalues of (a) and (b) respectively . (e) The 2D-plot of the flat band in the real part of frequencies at $c = \nu = 2$ for $\gamma = 2$. The presence of the imaginary part of the spectrum is well identified in (f).

## IV. RLC PARITY-TIME SYMMETRIC DIMER

### A. The RLC model and equations of dynamics

The RLC model of PT-Dimer is the one most investigated in

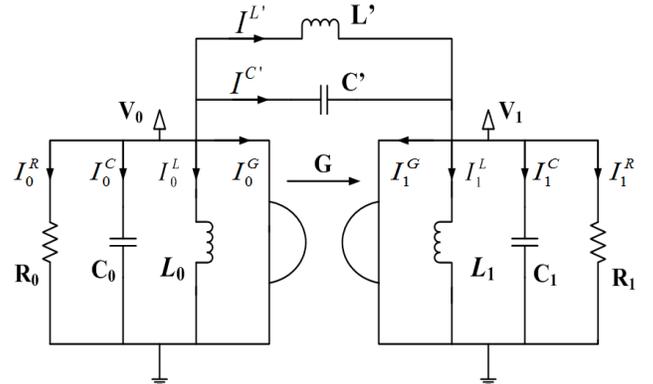

Fig. 6: The setup of the RLC dimer. The two RLC-cells are coupled by a capacitor $C'$, an inductor $L'$, a gyrator $G$ or by two (or by all) of them. Each cell contains a resistor $R_n$, a capacitor $C_n$ and an inductor $L'$.

Electronics up to now [7, 24, 25]. But it remains many interesting and intriguing phenomenon to discover in this classical oscillator. Our work is based on the investigation of coupled RLC active circuits by inductor, capacitor and gyrator. The Mutual coupling is left in order to simplified equations.

As described before in ZRC model, the non-hermiticity is assured by the coupling with the gyrator and gain/loss gather by positive and negative resistors in the setup. To avoid cells with imaginary natural frequencies, we assume that capacitor and inductors must automatically share the same sign to maintain the natural resonance frequency $\omega = \sqrt{1/LC}$ real. The setup is presented in Fig. 6 and the application of the Kirchoff's laws at nodes 0 and 1 give:

$$\begin{cases} I_0^R + I_0^L + I_0^C + I_0^G + I^{C'} + I^{L'} = 0 \\ I_1^R + I_1^L + I_1^C + I_1^G - I^{C'} - I^{L'} = 0 \end{cases} \quad (15)$$

All the currents are defined as in (7) except the currents in inductors expressed in $V_n = L_n \dfrac{d}{dt} I_n^L$ and $V_0 - V_1 = L' \dfrac{d}{dt} I^{L'}$.

These considerations lead on a general form of second order ordinary differential equations:

$$\begin{cases} (1+c_0)\dfrac{d^2}{d\tau^2}V_0 - c_0 \dfrac{d^2}{d\tau^2}V_1 = -(1+v_0)V_0 + v_0 V_1 - \gamma_0 \dfrac{dV_0}{d\tau} - g_0 \dfrac{dV_1}{d\tau} \\ -c_1 \dfrac{d^2}{d\tau^2}V_0 + (1+c_1)\dfrac{d^2}{d\tau^2}V_1 = \Gamma\left[v_1 V_0 - (1+v_1)V_1 - g_1 \dfrac{dV_0}{d\tau} - \gamma_1 \dfrac{dV_1}{d\tau}\right] \end{cases} \quad (16)$$

where $\tau = \omega_0 t$; $\omega_n = \dfrac{1}{\sqrt{L_n C_n}}$; $\Gamma = \dfrac{\omega_1}{\omega_0}$; $c_n = \dfrac{C'}{C_n}$; $v_n = \dfrac{L_n}{L'}$;

$\gamma_n = \dfrac{1}{R_n}\sqrt{\dfrac{L_n}{C_n}}$; $g_n = G_n \sqrt{\dfrac{L_n}{C_n}}$ and $\Delta = 1 + \sum_{n=0}^{1} c_n$.

Equation (16) is PT- symmetric when the serval conditions depicted in (9) are satisfied with new parameters. The PT-symmetric system is given as:

$$\begin{cases} (1+c)\dfrac{d^2}{d\tau^2}V_0 - c\dfrac{d^2}{d\tau^2}V_1 = -(1+v)V_0 + vV_1 - \gamma \dfrac{dV_0}{d\tau} - g \dfrac{dV_1}{d\tau} \\ -c\dfrac{d^2}{d\tau^2}V_0 + (1+c)\dfrac{d^2}{d\tau^2}V_1 = vV_0 - (1+v)V_1 + g\dfrac{dV_0}{d\tau} + \gamma \dfrac{dV_1}{d\tau} \end{cases} \quad (17)$$

Then, the elements of the resulted $4 \times 4$ effective Hamiltonian $H_{eff} = \begin{pmatrix} H_{11} & H_{12} \\ H_{21} & H_{22} \end{pmatrix}$ are related to the one of (10) with $H_{nm}$ being complex combination of Pauli matrices such that :

$$\begin{cases} H_{11} = 0 \\ H_{12} = j\mathbb{I} \\ H_{21} = jh_0 \mathbb{I} + jh_x \sigma_x \\ H_{22} = h_y \sigma_y + h_z \sigma_z \end{cases} \quad (18)$$

The eigenfunctions have the form of (19).

$$|\Psi(\tau)\rangle = \left(V_0(\tau), V_1(\tau), \dfrac{d}{d\tau}V_0(\tau), \dfrac{d}{d\tau}V_1(\tau)\right)^T \quad (19)$$

The eigenvalues of the Hamiltonian are given by:

$$\omega_{1,2,3,4} = \pm\sqrt{\dfrac{-2h_0 + h_y^2 + h_z^2 \pm \sqrt{\left(2h_0 - h_y^2 - h_z^2\right)^2 - 4\left(h_0^2 - h_x^2\right)}}{2}} \quad (20)$$

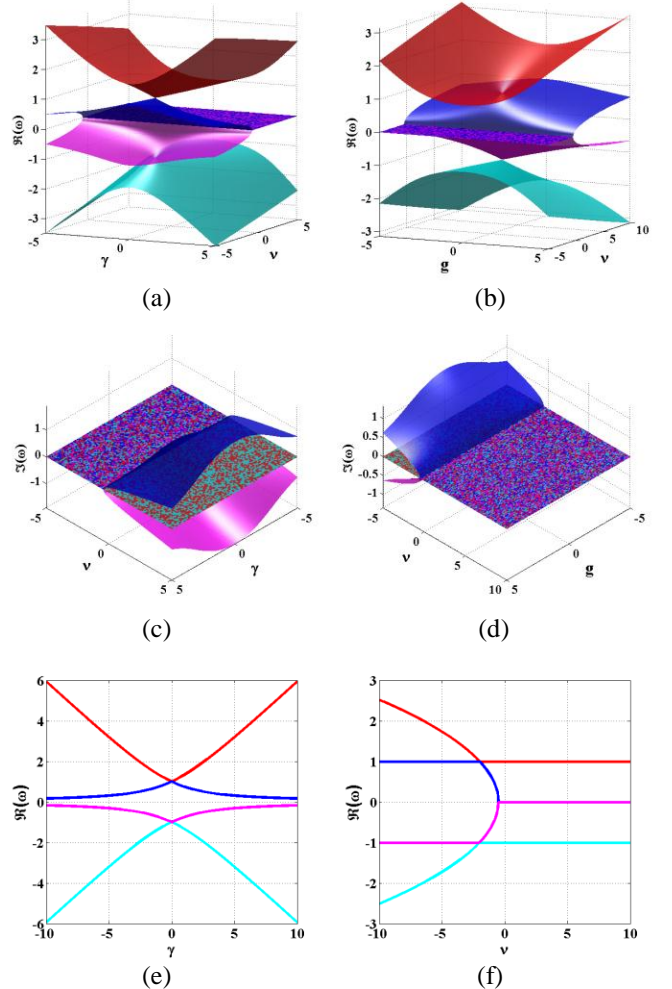

Fig. 7: (a) The double non-Hermitian diabolic points at $c=-2$ in unbreakable PT-symmetry induced by the gain and loss. (b) The double NHDP at $c=2$ in unbreakable PT-symmetry induced by gyroscopic coupling. (c) and (d) the imaginary parts of the eigenfrequencies of (a) and (b) respectively. The 2D-plot of the NHDP at $g=0$ function of (e) the gain/loss and (f) the imaginary coupling.

### B. Hermitian and non-Hermitian diabolic points and Ring of exceptional points in RLC-Dimer

*Hermitian diabolic points in PT-symmetric RLC-dimer*
The exploitation of the properties of the Hamiltonian give rise to the occurrence ounce more of the diabolic point at $c=v=-1$ without the gyroscopic coupling $g=0$. The setup turns into a Hermitian system although the presence of gain and loss in the dimer. In RLC-Dimer, the diabolic point occurs in pairs because in this second order oscillator, the frequencies are four contrary to ZRC-Dimer.

*Non-Hermitian diabolic points in PT-symmetric RLC-Dimer*
The non-Hermitian diabolic points occur in pairs (Fig. 7) in the unbreakable regions of parity time symmetry as shown in Table 1 for the condition $c = \nu$. Classically, in RLC PT-dimer, one has a transition from exact phase where eigenvalues are real to a complex phase and after the second transition all the eigenvalues are purely imaginary. The suppression of the complex phase attends on an exceptional point of order four (EP4) with a specific arrangement of the parameters of the system. In our case, the unbreakable PT-symmetry with NHDPs is presented to the best of our knowledge for the first time in Electronics with gyrators and negative capacitors or inductors. As we can see in the 2D-plot in Fig.7 e, the coalescence of eigenvalues happens once a time in a specific set of parameters. The occurrence of the NHDPs is robust to a Hermitian permutation as in ZRC-dimer in the conditions of unbreakable PT-symmetry when $c = \nu$ in the space of parameters.

*Ring of exceptional points in PT-symmetric RLC-Dimer*
Naturally, when the gain/loss or the gyroscopic coupling is added in a setup having a Hermitian or non-Hermitian diabolic point, the REP takes place in the spectrum (Fig.8). The REPs follow the curves of the of the equilateral hyperbola influenced in this case by the values of the imaginary coupling $\nu$ as given in (21)

$$\gamma^2 - g^2 = 2\left(1 + \nu + c \pm \sqrt{(1+2c)(1+2\nu)}\right) \quad (21)$$

## V. CONCLUSION

In this paper, we have demonstrated how to exhibit diabolic points, rings of exceptional points using the behavior of linear unbreakable PT-symmetry in ZRC and RLC-dimers. These setups may have many applications as in the design of novel antennas for high generation networks; in sensing and telemetry ; in the study of quantum interferences, Fano resonances, EIT, ATS and others related phenomena in positive and negative frequencies; in the fabrication of new optoelectronics devices and transistors like *WEYLFET*; in the mimic of the behavior of cold atoms at room temperature for significant exploitation in the quantum computer circuits; in the study of surface of exceptional points in topological material and semimetals by combining the two models (ZRC and RLC) in to order to increase the number of independent parameters [38, 60].

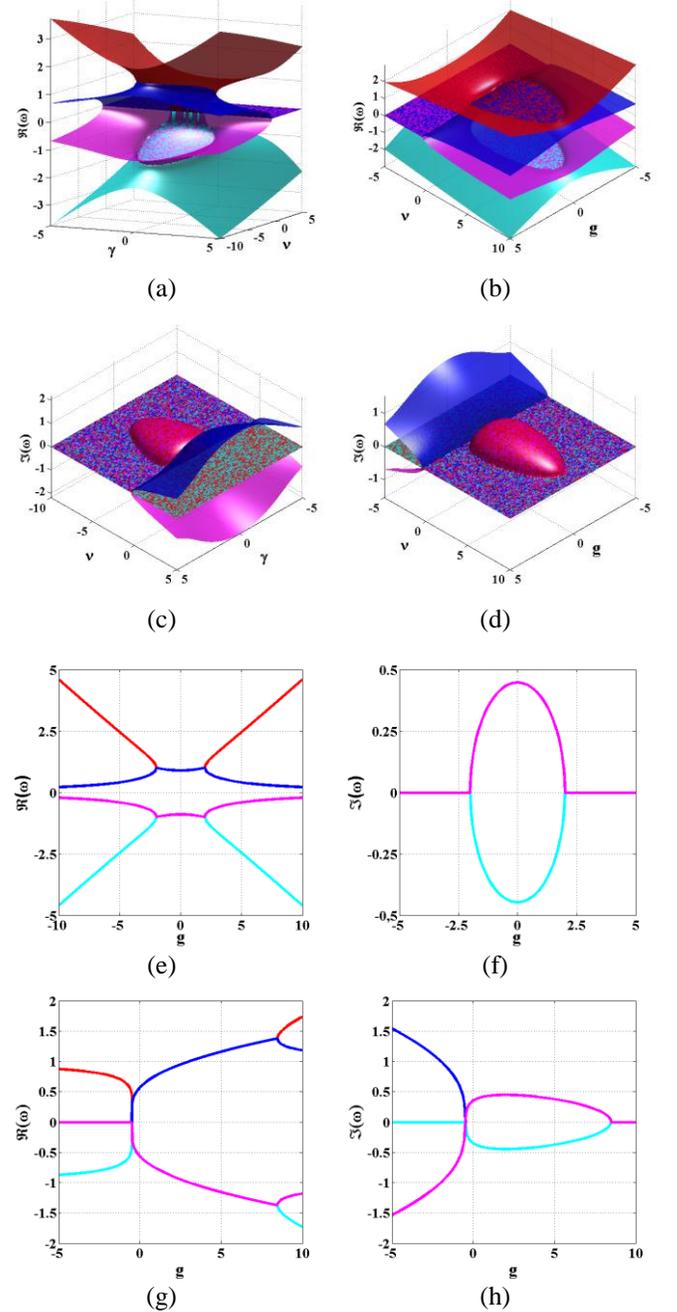

Fig. 8: (a) Double rings of exceptional points at $c = -2$ in unbreakable PT-symmetry induced by the gain and loss. (b) Double rings of exceptional points at $c = 2$ in unbreakable PT-symmetry induced by gyroscopic. (c) and (d) The imaginary of the eigenvalues of (a) and (b) respectively . (e) and (f) The 2D-plot of the real and imaginary parts of eigenfrequencies at $c = \nu = 2$ for $\gamma = 2$. (g) and (h) The 2D-plot of the real and imaginary parts of eigenfrequencies at $c = 2$ for $\gamma = 2$ function of the imaginary coupling. The particular case of $\nu = -1/2$ is well viewed in the 2D-plot of the eigenvalues with an exceptional point of order four (EP4).

## REFERENCES

[1] C. M. Bender and S. Boettcher, "Real Spectra in Non-Hermitian Hamiltonians Having PT-Symmetry," Physical Review Letters, vol. 80, no. 24, pp. 5243–5246, Jun. 1998.
[2] C. M. Bender, S. Boettcher, and P. N. Meisinger, "PT-symmetric quantum mechanics," Journal of Mathematical Physics, vol. 40, no. 5, pp. 2201–2229, May 1999.
[3] C. M. Bender, M. V. Berry, and A. Mandilara, "Generalized PT symmetry and real spectra," Journal of Physics A: Mathematical and General, vol. 35, no. 31, pp. L467–L471, Jul. 2002.
[5] S. Feng, "Loss-induced super scattering and gain-induced absorption," Optics Express, vol. 24, no. 2, p. 1291, Jan. 2016.
[6] S. Nixon and J. Yang, "All-real spectra in optical systems with arbitrary gain-and-loss distributions," Physical Review A, vol. 93, no. 3, Mar. 2016.

[7] F. Fotsa-Ngaffo, S. B. Tabeu, S. Tagouegni, and A. Kenfack-Jiotsa, "Thresholdless characterization in space and time reflection symmetry electronic dimers," Journal of the Optical Society of America B, vol. 34, no. 3, p. 658, Feb. 2017.

[8] S. B. Tabeu, F. Fotsa-Ngaffo, and A. Kenfack-Jiotsa, "Non-Hermitian Hamiltonian of two-level systems in complex quaternionic space: An introduction in electronics," EPL (Europhysics Letters), vol. 125, no. 2, p. 24002, Feb. 2019.

[9] A. Mostafazadeh, "Pseudo-Hermiticity versus PT symmetry: The necessary condition for the reality of the spectrum of a non-Hermitian Hamiltonian," Journal of Mathematical Physics, vol. 43, no. 1, pp. 205–214, Jan. 2002

[13] S. V. Suchkov, F. Fotsa-Ngaffo, A. Kenfack-Jiotsa, A. D. Tikeng, T. C. Kofane, Y. S. Kivshar, and A. A. Sukhorukov, "Non-Hermitian trimers: PT-symmetry versus pseudo-Hermiticity," New Journal of Physics, vol. 18, no. 6, p. 065005, Jun. 2016.

[14] A. Mock, "Comprehensive understanding of parity-time transitions in PT-symmetric photonic crystals with an antiunitary group theory," Physical Review A, vol. 95, no. 4, Apr. 2017.

[15] V. Lutsky, E. Luz, E. Granot, and B. A. Malomed, "Making the P T-Symmetry Unbreakable," Parity-time Symmetry and Its Applications, pp. 443–464, 2018.

[16] S. B. Tabeu, F. Fotsa-Ngaffo, and A. Kenfack-Jiotsa, "Imaginary resistor based Parity-Time symmetry electronics dimers," Optical and Quantum Electronics, vol. 51, no. 10, p. 335, Oct. 2019.

[17] C. Zheng, "Duality quantum simulation of a general parity-time-symmetric two-level system," EPL (Europhysics Letters), vol. 123, no. 4, p. 40002, Sep. 2018.

[18] R. El-Ganainy, K. G. Makris, M. Khajavikhan, Z. H. Musslimani, S. Rotter, and D. N. Christodoulides, "Non-Hermitian physics and PT symmetry," Nature Physics, vol. 14, no. 1, pp. 11–19, Jan. 2018.

[19] H. Zhao and L. Feng, "Parity–time symmetric photonics," National Science Review, vol. 5, no. 2, pp. 183–199, Jan. 2018.

[20] M.-A. Miri and A. Alù, "Exceptional points in optics and photonics," Science, vol. 363, no. 6422, p. eaar7709, Jan. 2019.

[21] Ş. K. Özdemir, S. Rotter, F. Nori, and L. Yang, "Parity–time symmetry and exceptional points in photonics," Nature Materials, vol. 18, no. 8, pp. 783–798, Apr. 2019.

[22] V. Lutsky, E. Luz, E. Granot, and B. A. Malomed, "Making the P T-Symmetry Unbreakable," Parity-time Symmetry and Its Applications, pp. 443–464, 2018.

[23] F. Klauck, L. Teuber, M. Ornigotti, M. Heinrich, S. Scheel, and A. Szameit, "Observation of PT-symmetric quantum interference," Nature Photonics, Sep. 2019.

[24] J. Schindler, Z. Lin, J. M. Lee, H. Ramezani, F. M. Ellis, and T. Kottos, PT-symmetric electronics," Journal of Physics A: Mathematical and Theoretical, vol. 45, no. 44, p. 444029, Oct. 2012.

[25] P.-Y. Chen, M. Sakhdari, M. Hajizadegan, Q. Cui, M. M.-C. Cheng, R. El-Ganainy, and A. Alù, "Generalized parity–time symmetry condition for enhanced sensor telemetry," Nature Electronics, vol. 1, no. 5, pp. 297–304, May 2018.

[26] N. Engheta, A. Salandrino, and A. Alù, "Circuit Elements at Optical Frequencies: Nanoinductors, Nanocapacitors, and Nanoresistors," Physical Review Letters, vol. 95, no. 9, Aug. 2005. Art. no. 095504

[27] P. Tassin, L. Zhang, T. Koschny, E. N. Economou, and C. M. Soukoulis, "Low-Loss Metamaterials Based on Classical Electromagnetically Induced Transparency," Physical Review Letters, vol. 102, no. 5, Feb. 2009.

[28] A. Alù and N. Engheta, "All Optical Metamaterial Circuit Board at the Nanoscale," Physical Review Letters, vol. 103, no. 14, Sep. 2009.

[29] H. Alaeian and J. A. Dionne, "Parity-time-symmetric plasmonic metamaterials," Physical Review A, vol. 89, no. 3, Mar. 2014.

[30] A. Baev, P. N. Prasad, H. Ågren, M. Samoć, and M. Wegener, "Metaphotonics: An emerging field with opportunities and challenges," Physics Reports, vol. 594, pp. 1–60, Sep. 2015.

[31] A. Cerjan, M. Xiao, L. Yuan, and S. Fan, "Effects of non-Hermitian perturbations on Weyl Hamiltonians with arbitrary topological charges," Physical Review B, vol. 97, no. 7, Feb. 2018. Art. no. 075128

[32] Y. Xu and C. Zhang, "Dirac and Weyl rings in three-dimensional cold-atom optical lattices," Physical Review A, vol. 93, no. 6, Jun. 2016.

[33] L.-K. Lim, J.-N. Fuchs, F. Piéchon, and G. Montambaux, "Dirac points emerging from flat bands in Lieb-kagome lattices," Physical Review B, vol. 101, no. 4, Jan. 2020. Art. no. 045131

[34] A. Zyuzin and P. Simon, "Disorder-induced exceptional points and nodal lines in Dirac superconductors," Physical Review B, vol. 99, no. 16, Apr. 2019. Art. no. 165145

[35] Z. Lin, A. Pick, M. Lončar, and A. W. Rodriguez, "Enhanced Spontaneous Emission at Third-Order Dirac Exceptional Points in Inverse-Designed Photonic Crystals," Physical Review Letters, vol. 117, no. 10, Aug. 2016.

[36] Q. Lin, M. Xiao, L. Yuan, and S. Fan, "Photonic Weyl point in a two-dimensional resonator lattice with a synthetic frequency dimension," Nature Communications, vol. 7, no. 1, Dec. 2016.

[37] Y. Lu, N. Jia, L. Su, C. Owens, G. Juzeliūnas, D. I. Schuster, and J. Simon, "Probing the Berry curvature and Fermi arcs of a Weyl circuit," Physical Review B, vol. 99, no. 2, Jan. 2019. Art. no. 020302(R)

[38] X. Zhang, K. Ding, X. Zhou, J. Xu, and D. Jin, "Experimental Observation of an Exceptional Surface in Synthetic Dimensions with Magnon Polaritons," Physical Review Letters, vol. 123, no. 23, Dec. 2019.

[39] J. Noh, S. Huang, D. Leykam, Y. D. Chong, K. P. Chen, and M. C. Rechtsman, "Experimental observation of optical Weyl points and Fermi arc-like surface states," Nature Physics, vol. 13, no. 6, pp. 611–617, Mar. 2017.

[40] A. Cerjan, S. Huang, M. Wang, K. P. Chen, Y. Chong, and M. C. Rechtsman, "Experimental realization of a Weyl exceptional ring," Nature Photonics, vol. 13, no. 9, pp. 623–628, Jun. 2019.

[41] V. Arjona, M. N. Chernodub, and M. A. H. Vozmediano, "Fingerprints of the conformal anomaly in the thermoelectric transport in Dirac and Weyl semimetals," Physical Review B, vol. 99, no. 23, Jun. 2019.

[42] A. A. Zyuzin and A. Y. Zyuzin, "Flat band in disorder-driven non-Hermitian Weyl semimetals," Physical Review B, vol. 97, no. 4, Jan. 2018.

[43] B. Yang, Q. Guo, B. Tremain, R. Liu, L. E. Barr, Q. Yan, W. Gao, H. Liu, Y. Xiang, J. Chen, C. Fang, A. Hibbins, L. Lu, and S. Zhang, "Ideal Weyl points and helicoid surface states in artificial photonic crystal structures," Science, vol. 359, no. 6379, pp. 1013–1016, Jan. 2018.

[44] C.-R. Mann, T. J. Sturges, G. Weick, W. L. Barnes, and E. Mariani, "Manipulating type-I and type-II Dirac polaritons in cavity-embedded honeycomb metasurfaces," Nature Communications, vol. 9, no. 1, Jun. 2018.

[45] S. Imhof, C. Berger, F. Bayer, J. Brehm, L. W. Molenkamp, T. Kiessling, F. Schindler, C. H. Lee, M. Greiter, T. Neupert, and R. Thomale, "Topolectrical-circuit realization of topological corner modes," Nature Physics, vol. 14, no. 9, pp. 925–929, Sep. 2018.

[46] W. B. Rui, M. M. Hirschmann, and A. P. Schnyder, "PT-symmetric non-Hermitian Dirac semimetals," Physical Review B, vol. 100, no. 24, Dec. 2019. Art. No. 245116

[47] I. Belopolski, P. Yu, D. S. Sanchez, Y. Ishida, T.-R. Chang, S. S. Zhang, S.-Y. Xu, H. Zheng, G. Chang, G. Bian, H.-T. Jeng, T. Kondo, H. Lin, Z. Liu, S. Shin, and M. Z. Hasan, "Signatures of a time-reversal symmetric Weyl semimetal with only four Weyl points," Nature Communications, vol. 8, no. 1, Oct. 2017.

[48] B. Zhen, C. W. Hsu, Y. Igarashi, L. Lu, I. Kaminer, A. Pick, S.-L. Chua, J. D. Joannopoulos, and M. Soljačić, "Spawning rings of exceptional points out of Dirac cones," Nature, vol. 525, no. 7569, pp. 354–358, Sep. 2015.

[49] Y. Zhang, Y. Sun, and B. Yan, "Berry curvature dipole in Weyl semimetal materials: An ab initio study," Physical Review B, vol. 97, no. 4, Jan. 2018.

[50] W.-Y. He and C. T. Chan, "The Emergence of Dirac points in Photonic Crystals with Mirror Symmetry," Scientific Reports, vol. 5, no. 1, Feb. 2015.

[51] Z. Yan and Z. Wang, "Tunable Weyl Points in Periodically Driven Nodal Line Semimetals," Physical Review Letters, vol. 117, no. 8, Aug. 2016.

[52] H.-X. Wang, Y. Chen, Z. H. Hang, H.-Y. Kee, and J.-H. Jiang, "Type-II Dirac photons," npj Quantum Materials, vol. 2, no. 1, Sep. 2017.

[53] A. A. Soluyanov, D. Gresch, Z. Wang, Q. Wu, M. Troyer, X. Dai, and B. A. Bernevig, "Type-II Weyl semimetals," Nature, vol. 527, no. 7579, pp. 495–498, Nov. 2015.

[54] M. Milićević, G. Montambaux, T. Ozawa, O. Jamadi, B. Real, I. Sagnes, A. Lemaître, L. Le Gratiet, A. Harouri, J. Bloch, and A. Amo, "Type-III and Tilted Dirac Cones Emerging from Flat Bands in Photonic Orbital Graphene," Physical Review X, vol. 9, no. 3, Jul. 2019. Art. No. 031010

[55] C.-L. Lin, R. Arafune, R.-Y. Liu, M. Yoshimura, B. Feng, K. Kawahara, Z. Ni, E. Minamitani, S. Watanabe, Y. Shi, M. Kawai, T.-C. Chiang, I. Matsuda, and N. Takagi, "Visualizing Type-II Weyl Points in Tungsten Ditelluride by Quasiparticle Interference," ACS Nano, vol. 11, no. 11, pp. 11459–11465, Oct. 2017.

[56] N. P. Armitage, E. J. Mele, and A. Vishwanath, "Weyl and Dirac semimetals in three-dimensional solids," Reviews of Modern Physics, vol. 90, no. 1, Jan. 2018.

[57] T. Ozawa, H. M. Price, A. Amo, N. Goldman, M. Hafezi, L. Lu, M. C. Rechtsman, D. Schuster, J. Simon, O. Zilberberg, and I. Carusotto, "Topological photonics," Reviews of Modern Physics, vol. 91, no. 1, Mar. 2019.

[59] Y. Xu, S.-T. Wang, and L.-M. Duan, "Weyl Exceptional Rings in a Three-Dimensional Dissipative Cold Atomic Gas," Physical Review Letters, vol. 118, no. 4, Jan. 2017. Art. No. 045701

[59] L. Lu, L. Fu, J. D. Joannopoulos, and M. Soljačić, "Weyl points and line nodes in gyroid photonic crystals," Nature Photonics, vol. 7, no. 4, pp. 294–299, Mar. 2013.

[60] S. B. Tabeu, F. Fotsa-Ngaffo, A. Kenfack-Jiotsa and Kazuhiro Shouno "Classification of positive, negative and imaginary components in non-hermitian systems and applications ," unpublished 2019.